\pdfoutput=1
\documentclass[aps,prl,reprint,superscriptaddress,longbibliography,floatfix]{revtex4-2}
\ifdefined\pdfsuppressptexinfo\pdfsuppressptexinfo=15\fi
\usepackage[T1]{fontenc}
\usepackage{amsmath,amssymb,bm,graphicx,booktabs,microtype}
\usepackage[colorlinks=true,allcolors=blue]{hyperref}
\hypersetup{pdftitle={Multipolar Nanophotonics with quantum computers},pdfauthor={Alex Krasnok}}
\newcommand{\ket}[1]{\lvert #1\rangle}
\newcommand{\bra}[1]{\langle #1\rvert}

\begin{document}
\title{Multipolar Nanophotonics with quantum computers}
\author{Alex Krasnok}
\affiliation{Department of Electrical and Computer Engineering, Florida International University, Miami, Florida 33174, USA}
\begin{abstract}
Small particles respond to light through electric and magnetic patterns called multipoles. Their interference determines the direction of scattered light and affects the decay of nearby emitters. Larger nanoantenna arrays, metasurfaces, and other nanophotonic systems demand increasingly costly simulations to meet strict performance and reliability requirements. We show how to represent these responses with quantum bits (qubits) and use quantum circuits to calculate scattering and the Purcell effect, the change in an emitter's decay rate caused by its surroundings. The circuits can encode known multipoles or estimate unknown ones from Maxwell's equations at the particle surface. We test the method against classical theory using an ideal quantum simulator and real IBM and IonQ quantum computers. For a silicon nanoparticle of radius $78$ nm illuminated at $679$ nm, IonQ gives a forward directivity of $2.984$, close to the ideal dipolar value $3$, quantifying the concentration of scattered light in the forward direction. For an emitter $10$ nm from a gold nanoparticle and emitting at $510$ nm, IBM measurements yield a calculated total decay rate $20.48$ times its vacuum value versus $20.47$ from theory, with correction for errors in reading the qubit states. We also establish how the number of multipoles and the required accuracy set the demands on qubits and measurements. These results motivate quantum-assisted design studies of directional radiation and emitter decay in nanophotonic systems.
\end{abstract}
\maketitle

\section*{Introduction}

Electric and magnetic multipoles provide a common language for electromagnetic systems, including antennas, resonators, metamaterials, and metasurfaces. They resolve sources and fields into components of distinct angular symmetry, describing radiation, scattering, absorption, and coupling through their amplitudes and phases. This description underlies Mie scattering \cite{mie1908}, antenna theory \cite{chu1948}, and transition-matrix methods \cite{waterman1971}. In nanophotonics, dipole interference produces directional scattering \cite{kerker1983,krasnok2012nano,kuznetsov2012,evlyukhin2012,fu2013} and supports optical Huygens elements \cite{krasnok2011huygens} and metasurfaces \cite{decker2015}. Higher magnetic multipoles enable superdirectivity and beam steering \cite{krasnok2014}, while the multipolar response near an emitter controls radiative enhancement, inhibition, and absorption \cite{krasnok2015}.

The coupling of these multipoles to radiation also determines how resonances confine energy. In an anapole response, electric and toroidal contributions cancel within one electric radiation channel, while other channels may remain open \cite{miroshnichenko2015}. In a periodic structure, the unit-cell multipoles couple to open diffraction channels. Symmetry or destructive interference can suppress every such coupling, producing a bound state in the continuum (BIC) in an ideal lossless system \cite{sadrieva2019}. Symmetry breaking creates radiative quasi-BICs \cite{koshelev2018}; interference in an isolated resonator can instead produce a finite-linewidth supercavity mode \cite{rybin2017}. These distinct mechanisms share a description in terms of complex multipole amplitudes and their coupling to allowed radiation outputs.

To predict these responses, multipole expansions represent the fields of spherical and nonspherical current distributions \cite{alaee2018}. Transition-matrix methods couple individual-particle responses through wave translations in clusters and periodic structures \cite{beutel2024}. This approach has enabled the design of an accidental BIC at a prescribed frequency and incidence angle \cite{gladyshev2023}. For large nonuniform assemblies, however, many multipoles and excitations must be solved together. Full-wave calculations can involve millions of unknowns, motivating methods that recover selected scattering outputs without reconstructing every field \cite{lin2022}. Fabrication and material variations require additional calculations to assess design reliability \cite{baranov2017}. A multipolar basis organizes this computation by physical contributions; its required size depends on the geometry and observable.

A quantum description of multipolar nanophotonics can retain the same physical design variables. The $2^q$ basis states of $q$ qubits label particles and electric or magnetic multipoles; gates transform their complex amplitudes, and detector projections estimate optical observables. The central problem is to connect this normalized register to the electromagnetic operator, excitation, and detector. Absolute scattering and Purcell factors also require the field norm and its phase relative to the source.

Quantum algorithms address electromagnetic scattering cross sections \cite{clader2013} and Maxwell-field evolution \cite{jin2024}, building on quantum linear-system methods \cite{hhl2009}. Related hardware demonstrations target other optical quantities. Vetlugin \emph{et al.}\ mapped single-photon interference and absorption onto a superconducting transmon \cite{vetlugin2022}. Fan \emph{et al.}\ simulated particle motion under optical forces using a force matrix from classical multipolar calculations \cite{fan2025}.

In this work we develop and test a multipolar formulation of quantum circuits for directional scattering and the Purcell effect. We connect electric and magnetic multipoles to register amplitudes, detector gates, and source-referenced optical quantities, using established methods for variational linear solving \cite{bravoprieto2023} and unnormalized-observable estimation \cite{alase2022}. We first encode known Mie coefficients to test preparation and interference, then estimate unknown coefficients from Maxwell boundary equations. Restoring normalization and source phase makes these coefficients usable for absolute scattering and radiative or absorptive emitter decay. A silicon sphere tests Huygens directionality; an emitter near silicon and gold tests radiative enhancement, inhibition, and absorption. Calculations (ideal quantum simulator) and measurements (IBM and IonQ quantum computers) distinguish preparation of a prescribed response from measured-feedback boundary recovery and show why a correct forward lobe can coexist with inaccurate total scattering or weak backward radiation. We also formulate the coupled-multipole extension and derive observable-specific accuracy and measurement requirements. Coupled-array and BIC calculations remain prospective.

\section*{Encoding known multipolar responses}

\begin{figure*}[!tp]
 \includegraphics[width=\textwidth]{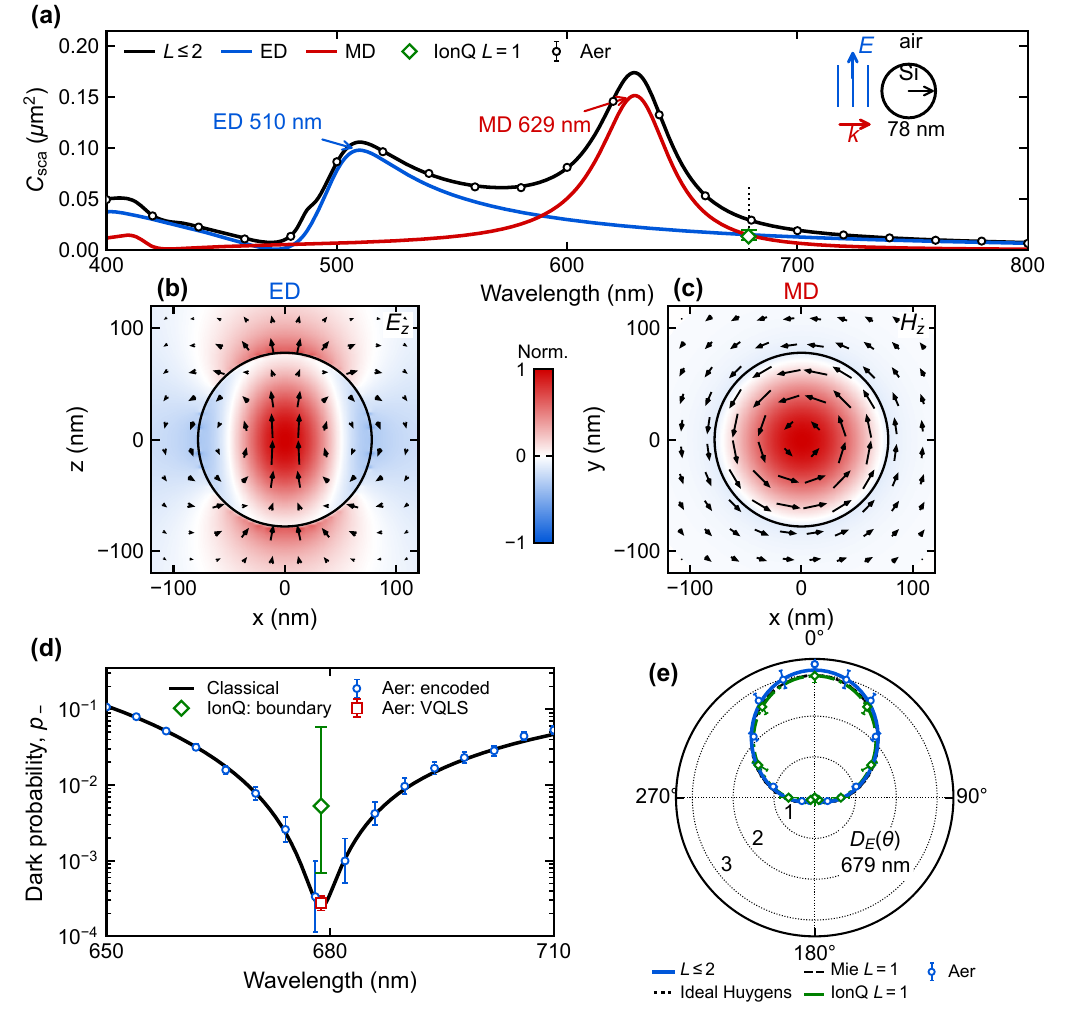}
 \caption{Multipolar scattering and Huygens interference for a silicon sphere of radius $78$ nm in air. (a) Classical electric-dipole (ED), magnetic-dipole (MD), and total $L\leq2$ cross sections; circles use Mie-loaded circuits (ideal quantum simulator, $65\,536$ shots). ``Aer: encoded'' loads classical Mie coefficients into gate angles; ``Aer: VQLS'' uses a variational quantum linear solver that adjusts those angles to satisfy the boundary equations. Labels mark partial-scattering maxima. The green diamond gives the raw boundary result (IonQ trapped-ion quantum computer), $C_{\rm sca}=0.0133\,[0.0097,0.0194]\,\mu\mathrm m^2$, versus the matched $L=1$ reference $0.0295\,\mu\mathrm m^2$. (b,c) Source-free dipole modes (classical Meep calculation). Color shows $\operatorname{Re}E_z$ in the ED meridional plane and $\operatorname{Re}H_z$ in the MD equatorial plane, normalized to the largest component magnitude inside the sphere. Arrows show $\operatorname{Re}\mathbf E$ for ED and $\operatorname{Im}\mathbf E$ for MD. Spherical symmetry permits these mode orientations independently of (a). (d) Dipole dark probability from Mie-loaded circuits ($131\,072$ shots per point) and exact-probability boundary optimization followed by optical readout ($4\,194\,304$ shots), both on an ideal quantum simulator. The raw IonQ boundary estimate is $0.0053\,[0.00070,0.059]$, versus the classical $2.43\times10^{-4}$. (e) E-plane directivity at $679$ nm. Blue curves and circles retain $L=2$ (ideal quantum simulator, $65\,536$ shots per projection, classical normalization); the dotted curve is an ideal Huygens source. Green IonQ and dashed black matched-classical curves retain $L=1$, each normalized by its integrated scattering. IonQ optical values are classical contractions of measured coefficients; $300$ whole-shot resamples of $1024$ six-bit records give conditional $95\%$ intervals at fixed measured angles. Aer bars are pointwise Wilson intervals. Angular samples are correlated, and IonQ reconstruction is raw and unconstrained.}
 \label{fig:scattering}
\end{figure*}

For the scattering benchmark, we take a crystalline-silicon sphere of radius $R=78$ nm in air, with Green's refractive index at $300$ K \cite{green2008}. A plane wave propagates along $z$, with electric field along $x$. Electric-dipole (ED) and magnetic-dipole (MD) scattering peak near $510$ and $629$ nm [Fig.~\ref{fig:scattering}(a)]. Mie theory \cite{mie1908} provides the classical reference. We use time dependence $e^{-i\omega t}$ and the outgoing-wave convention defined in the Supplement \cite{supplement}.

To identify the fields represented by the register, we calculate dipole modes independently with Meep \cite{oskooi2010} (classical electromagnetic solver). Meep advances Maxwell's equations on a spatial grid; pulsed currents select each dipole family, and the subsequent free decay gives its complex resonance frequency and profile [Figs.~\ref{fig:scattering}(b,c)]. The ED and MD pole wavelengths are approximately $498$ and $631$ nm. The ED mode describes electric polarization across the sphere; the MD mode has a circulating electric field and magnetic moment. Radiation and absorption make these source-free frequencies complex, so their real parts need not coincide with driven scattering maxima.

We construct the gate sequences with Qiskit, a toolkit for specifying quantum circuits, and evaluate them with Qiskit Aer, a classical simulator that numerically propagates quantum-state amplitudes. In an ideal finite-shot simulation, a shot is one circuit execution whose measurement outcome is sampled from those probabilities, without device errors. ``Exact Aer'' instead evaluates the probabilities directly, without sampling fluctuations; numerical precision, optimizer tolerance, and multipole truncation still limit the optical result.

To construct a circuit from an optical model, first choose the excitation, retained multipoles, and measured observable. Assign each multipole a binary label: ED and MD use $\ket0$ and $\ket1$; adding electric and magnetic quadrupoles gives $\ket{00},\ket{01},\ket{10},\ket{11}$. For $M$ amplitudes, use $q=\lceil\log_2M\rceil$ qubits and set unused labels to zero. The normalized, optically weighted coefficients become state amplitudes, while their physical norm is stored separately. For the sphere, this gives
\begin{equation}
 \ket{\psi_L}=\frac{1}{\sqrt{\mathcal N_L}}
 \sum_{\ell=1}^L\sqrt{2\ell+1}
 [a_\ell\ket{E,\ell}+b_\ell\ket{M,\ell}],
 \label{eq:state}
\end{equation}
where $\mathcal N_L=\sum_\ell(2\ell+1)(|a_\ell|^2+|b_\ell|^2)$ and $a_\ell,b_\ell$ are electric and magnetic Mie coefficients. Each basis label denotes a multipolar field component, not a photon in the particle. For host wavenumber $k$, the scattering cross section is $C_{\mathrm{sca}}=2\pi\mathcal N_L/k^2$. Known coefficients determine preparation angles directly; unknown coefficients require a boundary-solver circuit before optical reconstruction. Interacting systems additionally require the coupled operator discussed below.

Absolute observables require this physical norm. For multipole amplitudes $\mathbf c$ and a Hermitian optical operator $\mathsf W_{\mathcal O}$ that includes its dimensional prefactor,
\begin{equation}
 \mathcal O=\mathbf c^\dagger\mathsf W_{\mathcal O}\mathbf c
 =\|\mathbf c\|^2
 \bra{\psi}\mathsf W_{\mathcal O}\ket{\psi},
 \qquad
 \ket{\psi}=\frac{\mathbf c}{\|\mathbf c\|}.
 \label{eq:observable}
\end{equation}
Integrated scattering is diagonal in an orthogonal outgoing basis, whereas a directional detector combines multipoles coherently. Equation~\eqref{eq:observable} applies unnormalized-observable estimation \cite{alase2022} to these electromagnetic quantities. The choice of $\mathsf W_{\mathcal O}$ specifies both the detector and the operations required for its quantum measurement.

To measure the scattering norm, we add an auxiliary flag qubit. We prepare incident weights $w_j$, rotate the flag by a multipole-dependent angle to load $|t_j|$, and apply the phase $\arg t_j$, where $t_j=a_\ell$ or $b_\ell$. With $W=\sum_jw_j$, the flag-$1$ probability is
\begin{equation}
 p_s=\frac{\sum_j w_j|t_j|^2}{W},
 \qquad C_{\mathrm{sca}}=\frac{2\pi W}{k^2}p_s.
 \label{eq:flag}
\end{equation}
Here $w_{E,\ell}=w_{M,\ell}=2\ell+1$ and $W=2L(L+2)$. Passivity gives $|t_j|\leq1$ for the sphere in air, so the loading angles are real. Conditioning on flag $=1$ yields Eq.~\eqref{eq:state}. The other flag outcome completes the unitary encoding; it does not represent absorbed power.

Extinction requires a phase reference in addition to the norm:
\begin{align}
 C_{\mathrm{ext}}&=\frac{2\pi}{k^2}
 \sum_{\ell=1}^L(2\ell+1)\operatorname{Re}(a_\ell+b_\ell),
 \nonumber\\
 C_{\mathrm{abs}}&=C_{\mathrm{ext}}-C_{\mathrm{sca}}.
 \label{eq:extinction}
\end{align}
The real parts express interference with the incident wave. A common phase applied to every scattered amplitude leaves the quantum state unchanged but changes extinction. Recovering extinction and absorption therefore requires the response relative to the source. The boundary equation below provides this reference.

For dipoles, the common $\sqrt3$ weight in Eq.~\eqref{eq:state} cancels, giving $\ket{\psi_1}=(a_1\ket0+b_1\ket1)/\sqrt{s_d}$ with $s_d=|a_1|^2+|b_1|^2$. This normalized state can be prepared from $\ket0$ by $R_y(\theta)$ followed by $P(\delta)$, where $\theta=2\operatorname{atan2}(|b_1|,|a_1|)$ and $\delta=\arg b_1-\arg a_1$. The rotation sets the amplitude ratio and the phase gate sets the relative phase.

\begin{figure*}[tp]
 \includegraphics[width=\textwidth]{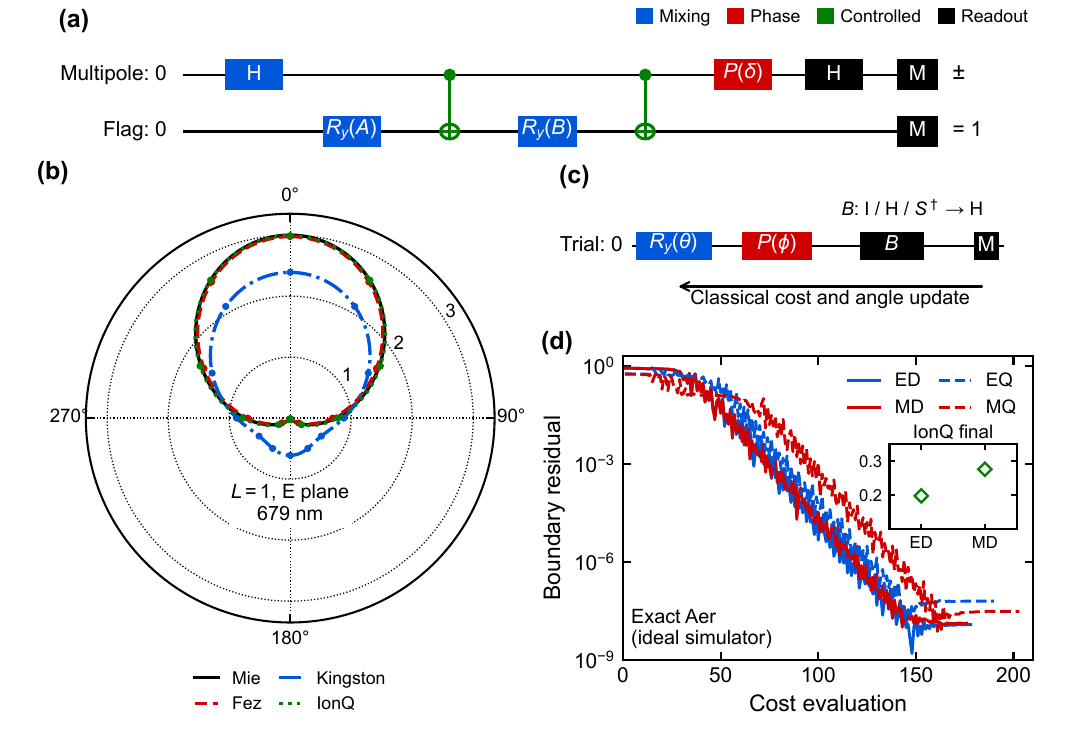}
 \caption{Quantum circuits and multipolar measurements. Gate colors identify amplitude mixing (blue), phase setting (red), controlled operations (green), and basis analysis or measurement (black). (a) Two-qubit dipole encoder (ideal quantum simulator). With $\alpha_E=\arcsin|a_1|$ and $\alpha_M=\arcsin|b_1|$, the flag rotations use $A=\alpha_E+\alpha_M$ and $B=\alpha_E-\alpha_M$; $P(\delta)$ sets $\delta=\arg b_1-\arg a_1$. A final $H$ reads sum and difference conditioned on flag $=1$. (b) Raw E-plane tomography, reconstructed from measurements in several bases, of Mie-loaded dipole states (real quantum computers): IBM Fez (red, $1024$ joint shots), IBM Kingston (blue, $4096$), and IonQ Forte Enterprise (green, $1024$ per packed circuit), against Mie theory (black). Conditional $95\%$ bars use $1000$ whole-shot resamples for IBM and $300$ for IonQ. Tomography is raw and unconstrained. These prepared-state measurements test neither the flagged encoder nor boundary recovery. (c) Variational boundary circuit. Here the basis-analysis gate $B$ is $I$ for $Z$, $H$ for $X$, or $S^\dagger$ followed by $H$ for $Y$. (d) Physical residuals near $679$ nm for ED, MD, electric-quadrupole (EQ), and magnetic-quadrupole (MQ) optimization (ideal quantum simulator, exact probabilities). The inset gives raw final residuals $0.20$ and $0.28$ after measured phase and polar-angle updates (IonQ quantum computer), on a separate linear scale.}
 \label{fig:circuits}
\end{figure*}

Figure~\ref{fig:circuits}(a) adds a flag to measure the scattering norm as well. The first Hadamard $H$ prepares equal ED/MD weights. A flag rotation $R_y(\gamma)$ gives amplitudes $\cos(\gamma/2)$ and $\sin(\gamma/2)$, while a controlled-NOT (CX) flips the flag when the multipole bit is $1$. Setting $\alpha_E=\arcsin|a_1|$ and $\alpha_M=\arcsin|b_1|$ fixes the two rotation angles to $A=\alpha_E+\alpha_M$ and $B=\alpha_E-\alpha_M$. The sequence $R_y(A)$, CX, $R_y(B)$, CX produces flag rotations $A+B=2\alpha_E$ for ED and $A-B=2\alpha_M$ for MD. Applying $P(\delta)$ then supplies their relative phase. Flag-$1$ selection recovers $\ket{\psi_1}$ with probability $p_s=s_d/2$; its frequency supplies the norm in Eq.~\eqref{eq:flag}.

The final $H$ implements the optical detector. Along the illumination axis, ED and MD have equal forward and opposite backward angular factors. Acting on $\ket{\psi_1}$, $H$ gives amplitudes $(a_1+b_1)/\sqrt{2s_d}$ for bit $0$ and $(a_1-b_1)/\sqrt{2s_d}$ for bit $1$. Squaring the latter gives the backward probability conditioned on flag $=1$:
\begin{equation}
 p_-=\frac{|a_1-b_1|^2}{2(|a_1|^2+|b_1|^2)}
 =\frac{1-\langle X\rangle}{2},
 \qquad
 \frac{I(0)}{I(\pi)}=\frac{1-p_-}{p_-},
 \label{eq:dark}
\end{equation}
where $X$ exchanges the electric and magnetic labels, so $\langle X\rangle=2\operatorname{Re}(a_1^*b_1)/s_d$. The measured sum and difference probabilities thus reproduce the axial intensity ratio. At the Huygens condition $a_1=b_1$, the state $\ket+=(\ket0+\ket1)/\sqrt2$ has a dark difference output. Its directivity is $D(\vartheta)=3(1+\cos\vartheta)^2/4$, with $D(0)=3$. A large axial forward/backward ratio therefore indicates a directional cancellation, without implying high directivity or weak total scattering.

For silicon, the dipole mismatch is smallest near $679$ nm [Fig.~\ref{fig:scattering}(d)], where $|b_1/a_1|=0.991$, $\arg(b_1/a_1)=1.7^\circ$, and $p_-=2.43\times10^{-4}$. The dipole cross section remains $0.0295\ \mu\mathrm m^2$, while the axial intensity ratio reaches approximately $4.1\times10^3$. Higher multipoles contribute only $0.0404\%$ of scattered power but change the forward directivity from approximately $3$ to $3.13$. Integrated-power convergence thus does not ensure convergence of an interference-sensitive pattern. We retain both quadrupoles in the polar calculation [Fig.~\ref{fig:scattering}(e)]. For linear polarization, the E- and H-plane cross sections are $|S_2|^2/k^2$ and $|S_1|^2/k^2$; they coincide for ideal dual dipoles, but spherical geometry alone does not make them equal.

In ``Aer: encoded'' or Mie-loaded calculations, classical Mie coefficients set the gate rotations and phases; Aer then evaluates their interference (ideal quantum simulator). These circuits test preparation and readout of a prescribed response. Exact probabilities reproduce it to numerical precision; finite shots add sampling uncertainty. Coherence and entanglement play distinct roles: unequal dipole magnitudes generally entangle the multipole and flag registers, whereas equal magnitudes give a separable final state at any relative phase. The normalized dipole pair itself needs one qubit. Huygens interference requires electric--magnetic coherence; CX gates implement conditional magnitude loading in this encoding.

To test preparation on hardware before boundary recovery, we directly load the normalized Mie dipole state near $679$ nm (IBM Fez quantum computer). Fez and Kingston are IBM processors whose superconducting transmon qubits are controlled by microwave signals; physical preparation and measurement introduce device errors as well as finite-count fluctuations. We reconstruct the state by tomography, combining measurements in the $X,Y,Z$ bases to determine its populations and coherence. Each of $1024$ joint shots contains six sequential measurement and control blocks, with $2$ s of reported QPU usage. The raw Pauli components $(0.984,0.057,0.074)$ give target overlap $0.993$ and forward directivity $2.98$, versus the classical value $\approx3$ [Fig.~\ref{fig:circuits}(b)]. Angular estimates follow from this state; resampling complete shots retains interblock correlations.

The dark direction is more sensitive to hardware error. Eight backward outcomes give $p_-^{\mathrm{raw}}=0.0078$, whereas the ideal probability predicts $0.25$ events in $1024$ shots. Ideal-Huygens and zero-state controls give $30$ and $9$ erroneous outcomes. Their different preparation circuits prevent separating preparation and readout errors uniquely. The raw data resolve the main lobe but not the deep predicted minimum: high state overlap can coexist with a large relative error in weak scattering.

We next extend prepared-state tomography through $L=1,2,3$ (IBM Kingston quantum computer), using $4096$ joint shots and $5$ s of QPU usage for tomography, controls, and initial boundary costs. Raw target overlaps are $0.793$, $0.566$, and $0.526$. The dipole estimates $p_-^{\mathrm{raw}}=0.204$ and $D(0)=2.39$ differ substantially from the ideal response [Fig.~\ref{fig:circuits}(b)]. Zero-state errors on two allocated qubits range from approximately $18\%$ to $29\%$ across monitor blocks. Processor and acquisition differences prevent attributing the poorer agreement solely to multipole order. The Supplement retains the higher-order patterns and slightly indefinite $L=3$ density matrix. The initial cost measurements supply a phase update for ED and MD, but this experiment does not complete boundary recovery.

We also prepare the same Mie-loaded states on IonQ Forte Enterprise 1 (real quantum computer). This trapped-ion processor stores qubits in atomic internal states, implements gates with laser-driven operations, and reads the states through fluorescence. Raw overlaps are $0.998$, $0.986$, and $0.942$ for $L=1,2,3$. Three dipole dark events in $1024$ shots give $p_-^{\mathrm{raw}}=0.00293$ and forward directivity $2.99$ [Fig.~\ref{fig:circuits}(b)]. Five science circuits and two full-register controls pack independent registers, each measured $1024$ times; IBM reuses qubits sequentially. These acquisition differences, unequal shot counts, and compiler mappings prevent isolating qubit technology as the cause of the contrast. IonQ tomography is raw and unconstrained, including the negative $L=3$ backward estimate in the Supplement. These experiments test multipolar state preparation from classical coefficients.

\section*{Recovering multipoles from boundary equations}

To calculate unknown multipole amplitudes, we use a variational quantum linear solver (VQLS) \cite{bravoprieto2023}: a trial circuit prepares adjustable amplitudes, and a classical optimizer changes its gate angles to minimize a boundary-equation residual. ``Aer: VQLS'' denotes this solver evaluated in Aer; exact-probability optimization supplies its costs without measurement sampling.

Tangential-field continuity gives a complex $2\times2$ system $A\mathbf x=\mathbf f$ for each electric or magnetic multipole's scattered and internal amplitudes. The matrix contains radial functions, geometry, and material dispersion. The circuit $\ket u=P(\phi)R_y(\theta)\ket0$ spans the normalized solution direction [Fig.~\ref{fig:circuits}(c)]. With $Q=I-\ket{\hat f}\bra{\hat f}$ and $\hat f=\mathbf f/\|\mathbf f\|$, the positive operator $A^\dagger QA$ penalizes the component of $A\ket u$ perpendicular to the excitation. Its expectation vanishes at the solution. Expanding it in $I,X,Y,Z$ gives three measurement bases: direct readout for $Z$, $H$ for $X$, and $S^\dagger=P(-\pi/2)$ followed by $H$ for $Y$. Measured expectations guide classical updates of the two angles. Thus the optical boundary matrix determines the measured cost, while the gates represent the unknown solution. This two-dimensional operator is constructed and readily diagonalized classically; the circuit tests boundary recovery, without providing a computational advantage for a sphere.

The optimized state specifies a direction but not the field amplitude. To restore its magnitude and source phase, we minimize the physical residual over a complex scale:
\begin{equation}
 s=\frac{(A\mathbf u)^\dagger\mathbf f}{\|A\mathbf u\|^2},
 \qquad \mathbf x=s\mathbf u.
 \label{eq:scale}
\end{equation}
The opposite phases of $\mathbf u$ and $s$ cancel, leaving $\mathbf x$ invariant. The prescribed excitation fixes the physical norm and source phase needed for extinction and the Purcell effect.

We apply this recovery to $20$ boundary systems (ideal quantum simulator): ED and MD at nine wavelengths, plus EQ and MQ at the Huygens wavelength. Optimization with exact Aer probabilities gives a maximum absolute coefficient error of $1.4\times10^{-8}$ and relative boundary residual of $6.4\times10^{-8}$ [Fig.~\ref{fig:circuits}(d)]. Transferring these coefficients classically to the interference circuit gives $77$ dark events among $281\,021$ successful trials in $4\,194\,304$ shots. The resulting $p_-=2.74\times10^{-4}$ has a $95\%$ interval $[2.2,3.4]\times10^{-4}$ containing the classical reference. This calculation links exact-probability boundary optimization to finite-shot optical readout.

To determine coefficients from measured boundary costs, we perform one phase update, one polar-angle update, and final tomography for ED and MD near $679$ nm (IonQ quantum computer). Each continuation circuit returns $1024$ packed shots, with raw feedback and reconstruction. Classical scale recovery gives $C_{\rm sca}=0.0133\,\mu\mathrm m^2$, with conditional $95\%$ interval $[0.0097,0.0194]\,\mu\mathrm m^2$, versus the matched dipole reference $0.0295\,\mu\mathrm m^2$ [Fig.~\ref{fig:scattering}(a)]. ED and MD residuals are $0.198$ and $0.277$ [Fig.~\ref{fig:circuits}(d)]. The coefficients give $D(0)=2.984$ and $p_-=0.00529$, versus $\approx3$ and $2.43\times10^{-4}$ classically [Fig.~\ref{fig:scattering}(d,e)]. The dark-probability interval $[0.00070,0.0586]$ excludes the reference. Thus the forward lobe remains nearly unchanged despite a $55\%$ underestimate of absolute scattering and excess backward radiation. These optical values are classical contractions of measured boundary amplitudes. Their uncertainty is conditional on the measured optimizer angles.

\section*{Purcell effect}

The recovered multipole coefficients also let us calculate the Purcell effect. We place a tangential dipole $\mathbf d=d\hat{\mathbf x}$ at $\mathbf r_0=(0,0,R+g)$, a surface gap $g$ from the sphere. In the weak-coupling, Markov regime, its electromagnetic decay rate relative to vacuum equals the normalized power supplied by the corresponding classical dipole \cite{chew1987,ruppin1982,krasnok2015}:
\begin{equation}
 F_{\mathrm{tot}}=1+\frac{6\pi}{k}
 \hat{\mathbf d}\cdot\operatorname{Im}\mathbf G_s(\mathbf r_0,\mathbf r_0)
 \cdot\hat{\mathbf d},
 \qquad F_{\mathrm{tot}}=F_{\mathrm{rad}}+F_{\mathrm{abs}}.
 \label{eq:purcell}
\end{equation}
Here $\mathbf G_s$ is the scattering electric Green tensor and $\hat{\mathbf d}$ the real unit dipole vector. Radiation and particle absorption contribute $F_{\mathrm{rad}}$ and $F_{\mathrm{abs}}$. The model excludes a substrate and intrinsic nonradiative emitter loss.

The directly emitted and induced fields contribute coherently to radiation:
\begin{equation}
 F_{\mathrm{rad}}=1+F_{\mathrm{sca}}+F_{\mathrm{int}},
 \label{eq:interference}
\end{equation}
where $F_{\mathrm{sca}}$ is the induced-field power and $F_{\mathrm{int}}$ its signed interference with the direct source field. A strong induced field can therefore inhibit radiation. Adding the direct-source amplitude to the response vector puts this power in the form of Eq.~\eqref{eq:observable}. We calculate the multipole coefficients with independent one-qubit boundary circuits, then contract them classically with the Green tensor and power operators. The qubits encode boundary amplitudes, without representing the microscopic quantum state of the emitter or material.

\begin{figure*}[tp]
 \includegraphics[width=\textwidth]{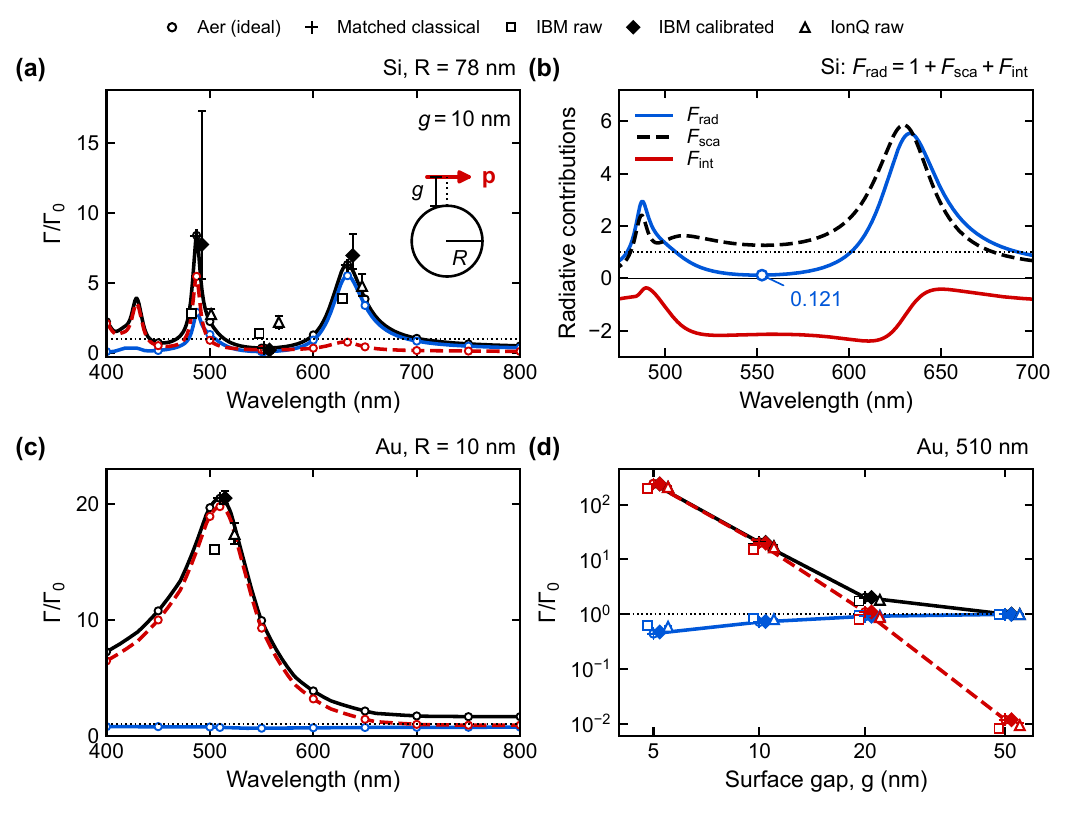}
 \caption{Purcell effect near dielectric and plasmonic spheres. A tangential dipole lies a gap $g$ from a sphere in air. (a) Si spectra ($R=78$ nm, $g=10$ nm), normalized to vacuum. (b) Si radiation decomposed into direct-source, induced-field, and interference contributions: radiative power is blue solid, induced-field power black dashed, and interference red solid. The circle uses the ideal quantum simulator near the $553$ nm total-rate minimum. (c) Au spectra ($R=10$ nm, $g=10$ nm), using Johnson--Christy film constants. (d) Au rates at $510$ nm and four gaps. In (a,c,d), total, radiative, and absorptive factors are black solid, blue solid, and red dashed. Spectral curves are converged classical sums; circles use exact-probability boundary coefficients (ideal quantum simulator) and classical optical contractions. Lines between simulator points in (d) guide the eye. Open squares and filled diamonds give raw and readout-calibrated IBM Kingston results; open triangles give raw IonQ Forte Enterprise 1 results (real quantum computers). Hardware markers show total factors in (a,c) and all three components in (d), using black, blue, and red for total, radiative, and absorptive rates. Plus signs give classical references with the same multipole truncation. Hardware bars are conditional $95\%$ intervals from $300$ whole-shot resamples at fixed measured optimizer angles, including readout calibration for IBM. IonQ uses raw feedback and reconstruction. Small horizontal offsets separate markers. Each platform solves the same $80$ independent one-qubit boundary systems, followed by classical optical reconstruction.}
 \label{fig:purcell}
\end{figure*}

For silicon at $g=10$ nm, the classical calculation gives a radiative maximum $F_{\mathrm{rad}}=5.54$ near $633$ nm [Fig.~\ref{fig:purcell}(a)]. The total factor $6.33$ corresponds to a radiative fraction of approximately $87.5\%$. Near the total-rate minimum at $553$ nm, $F_{\mathrm{tot}}=0.372$, $F_{\mathrm{rad}}=0.121$, and $F_{\mathrm{abs}}=0.251$. Although the induced-field power $F_{\mathrm{sca}}=1.26$ exceeds the vacuum source power, interference $F_{\mathrm{int}}=-2.14$ suppresses radiation [Fig.~\ref{fig:purcell}(b)]. At the total-rate maximum near $487$ nm, $F_{\mathrm{tot}}=8.42$ includes absorption $5.49$. A large dielectric Purcell factor can thus contain substantial dissipation.

Gold makes the dissipative contribution dominant [Fig.~\ref{fig:purcell}(c)]. For $R=10$ nm and $g=10$ nm, Johnson--Christy optical constants \cite{johnson1972} give a classical total-rate peak of $20.5$ near $510$ nm, with radiative factor $0.717$ and absorptive factor $19.8$. Only $3.5\%$ of the supplied power reaches the far field. Reducing the gap from $50$ to $5$ nm strongly increases absorption while radiation remains below its vacuum value [Fig.~\ref{fig:purcell}(d)]. This plasmonic Purcell enhancement therefore represents predominantly quenching. The local-response model uses measured film constants; fluorescence also depends on excitation and intrinsic quantum yield \cite{anger2006}. The Supplement tests size-dependent damping motivated by nanoparticle measurements \cite{berciaud2005}.

To resolve the Purcell effect, more multipoles are needed than for plane-wave scattering because a local source excites finer spatial variation. At $g=10$ nm, the first tested truncations with less than $1\%$ maximum total-rate error are $L=32$ for Si and $L=6$ for Au, relative to $L=128$ and $48$ over the respective wavelength grids. For Si at $g=5$ nm, extending the reference from $128$ to $512$ changes the total rate by at most $1.3\times10^{-5}$ relatively. These convergence tests provide no analytic remainder bound. Each electric or magnetic boundary system still needs one data qubit, while the number of solves grows with $L$. Selected spectral points use boundary coefficients from circuits (ideal quantum simulator); classical sums establish spectral convergence.

We then calculate selected Purcell factors from $80$ independent one-qubit boundary systems (IBM Kingston quantum computer). Three stages measure phase updates, polar-angle updates, and final Pauli expectations with $16\,384$, $16\,384$, and $65\,536$ shots per circuit. Local zero- and one-state measurements calibrate readout at every stage. The readout-calibrated costs determine subsequent angles; no solved Mie coefficients or simulator-optimal angles enter this feedback. Classical scale recovery and optical contractions give three silicon spectral points at $g=10$ nm and four gold gaps near $510$ nm. Each comparison retains exactly the same electric and magnetic orders as its classical reference. The stages use $49$, $49$, and $88$ s of billed QPU time, totaling $186$ s, or $193$ s including both prepared-state experiments.

The IBM coefficients resolve dielectric inhibition and plasmonic enhancement with different accuracies. For Si near $553$ nm, the raw and readout-calibrated total factors are $1.39$ and $0.237$. The latter has conditional $95\%$ interval $[0.038,0.403]$, compared with the matched classical value $0.369$. The calibrated radiative factor is $0.120\,[0.112,0.132]$, versus $0.121$. Near $633$ nm it is $5.99\,[4.30,9.12]$, versus $5.54$. For Au at $g=10$ nm, the raw total factor is $16.0$ and the calibrated factor $20.48\,[19.94,21.11]$, versus $20.47$. The absorbed factor $19.75$ retains the dissipative character of this Purcell enhancement.

Several discrepancies remain. The matched references fall below the calibrated intervals for gold's radiative factors at $g=5$, $10$, and $20$ nm and total factor at $g=50$ nm. At silicon's $487$ nm point, the calibrated total estimate $7.77$ has interval $[5.29,17.28]$. The Supplement reports all raw and calibrated rates, including negative estimates. The intervals resample final tomography and calibration at fixed measured optimizer angles; they exclude repeated-optimization variability, drift, gate bias, and material uncertainty. They therefore quantify conditional sampling uncertainty, not the full error of the hardware calculation.

We solve the same $80$ boundary systems on IonQ Forte Enterprise 1 (real quantum computer), using $2048$ shots per circuit for each optimization stage and $8192$ for final tomography. Feedback and reconstruction remain raw. For Au at $510$ nm and $g=10$ nm, the total factor is $17.39\,[16.55,18.34]$, $15\%$ below the matched $20.47$ [Fig.~\ref{fig:purcell}(c,d)]. At the Si inhibition point near $553$ nm, the total factor is $2.24\,[1.78,2.68]$ versus $0.369$ [Fig.~\ref{fig:purcell}(a)], although the radiative factor remains below unity, $0.188$ versus $0.121$. The absorbed contribution, $2.05$ versus $0.248$, accounts for most of this total-rate discrepancy. None of the $21$ raw IonQ rate intervals contains its matched reference. The Supplement compares all rates and gives an ideal-readout diagnostic at the measured optimizer angles.

\section*{Discussion}

The same multipole coefficients determine scattering and the Purcell effect through different excitations and optical contractions. Electric--magnetic interference controls Huygens directionality; direct--induced interference controls radiative enhancement and inhibition, with absorption adding dissipation.

Loading known coefficients tests preparation and readout; minimizing boundary costs estimates the response. The IonQ boundary result preserves the bright lobe despite substantial amplitude error, requiring separate validation of absolute and weak optical signals. Matrix construction and optical reconstruction remain classical. The $80$ parallel one-qubit solves contain no coupling between different multipole systems. Mie theory is more efficient for a sphere; potential computational gains concern interacting multipoles.

Hardware accuracy also depends on the physical qubits and the calibration state at execution. IBM Purcell circuits retain the same $80$ physical qubits across stages and measure local zero- and one-state controls within each stage. Final one-to-zero error fractions vary from $0.43\%$ to $12.93\%$ across these qubits, so one processor-wide readout correction would miss substantial variation. IonQ controls remain diagnostic because their physical-ion mapping to science circuits is unverified. The preserved records do not establish the providers' gate-calibration times; elapsed time since calibration and intervening drift therefore remain unresolved. Same-stage readout controls constrain assignment errors without calibrating gate errors or establishing long-term stability.

\begin{figure*}[tp]
 \includegraphics[width=\textwidth]{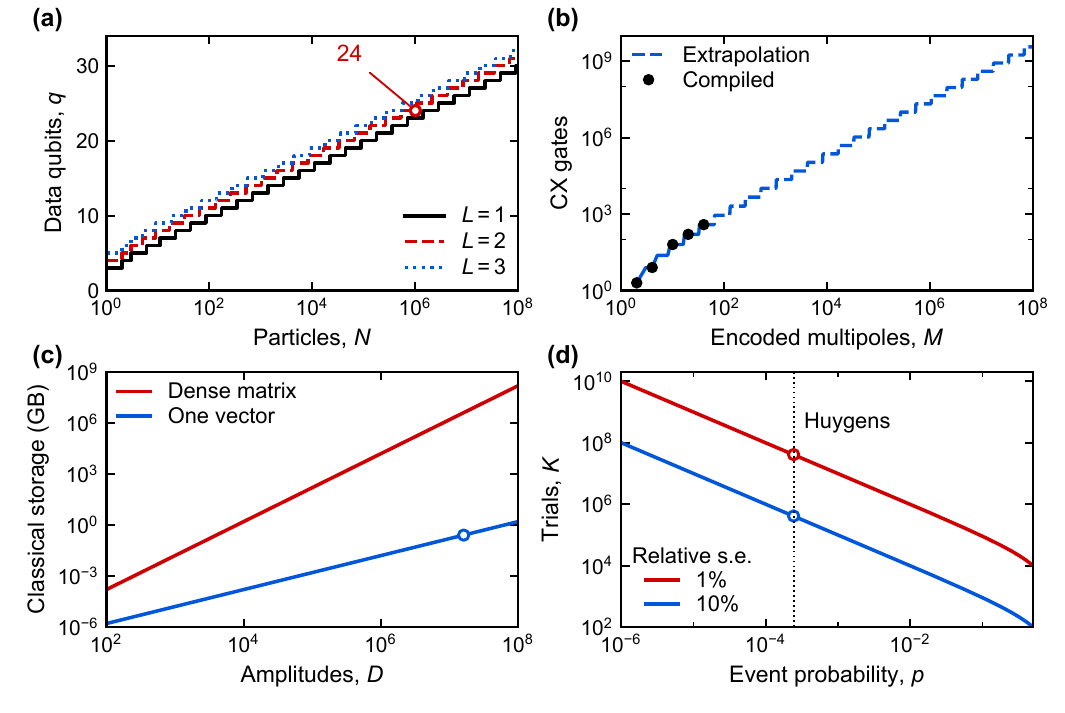}
 \caption{Register size and computational costs. (a) Data qubits for $N$ particles with electric and magnetic multipoles through $L$, using Eq.~\eqref{eq:arrays}. (b) CX count of the explicit generic encoder: compiled small-register circuits and algebraic extrapolation. (c) Complex-double storage of one response vector and a hypothetical dense interaction matrix. Efficient classical solvers can avoid dense storage. (d) Independent accepted trials for relative standard error $r$, including the silicon Huygens probability. These estimates do not establish an interacting-array solution or an advantage threshold.}
 \label{fig:scaling}
\end{figure*}

For interacting structures, the data register grows logarithmically with the number of amplitudes. The reduced sphere basis has $M=2L$ multipoles and $q=\lceil\log_2M\rceil$ qubits. Ten total multipoles need four data qubits; ten electric plus ten magnetic multipoles need five. The passive encoder adds one flag. General particle interactions require every azimuthal component, giving
\begin{equation}
 D=2N\sum_{\ell=1}^L(2\ell+1)=2NL(L+2),
 \qquad q=\lceil\log_2D\rceil.
 \label{eq:arrays}
\end{equation}
for $N$ particles and both multipole families through order $L$. Each particle contributes six dipole amplitudes, or sixteen through quadrupoles. A million particles therefore need $23$ or $24$ data qubits [Fig.~\ref{fig:scaling}(a)], before solver workspaces and error correction.

Loading this compact register requires a number of independent parameters proportional to the dimension for an arbitrary vector \cite{mottonen2005}. Our encoder uses $q2^q$ CX gates for preparation and $(2q-1)2^q$ with the axial detector, padding to $2^q$ states [Fig.~\ref{fig:scaling}(b)]. At $q=24$, preparation requires $402\,653\,184$ CX gates. Exploiting symmetry or analytic input structure can reduce these construction-specific counts; arbitrary input lists still require processing.

A predictive calculation must also solve the interactions. In the transition-matrix description \cite{beutel2024}, $\mathsf T$ contains individual-particle responses and $\mathsf U$ translates outgoing waves into regular waves at other particles. Their coupled amplitudes satisfy
\begin{equation}
 (I-\mathsf T\mathsf U)\mathbf c=\mathsf T\mathbf a_{\mathrm{inc}}.
 \label{eq:coupled}
\end{equation}
For nonspherical particles, each $\mathsf T$ block can mix orders and multipole families and requires an electromagnetic calculation. Sphere entries are $-a_\ell$ and $-b_\ell$ in the Supplement's convention. Quantum solution of Eq.~\eqref{eq:coupled} remains prospective. Favorable scaling requires suitable operator access and conditioning \cite{costa2022}, controlled accuracy, and efficient input and output \cite{montanaro2016}. Long-range translations generally make the matrix dense; collective resonances can worsen conditioning.

The required solve depends on the design objective. A superdirective nanoantenna must excite an appropriate combination of higher multipoles \cite{krasnok2014}; quantum readout would project onto the desired angular mode. A BIC calculation must find a self-consistent eigenmode of the periodic interaction operator and verify zero coupling to every open diffraction channel \cite{sadrieva2019,gladyshev2023}. Cancelling one backward Huygens output does not meet these conditions. Such extensions require coupled solves or eigenvalue calculations and their detector projections, beyond the one-qubit sphere demonstration.

Computational comparisons must preserve this electromagnetic structure. A complex-double vector with $1.6\times10^7$ entries occupies $256$ MB; its explicitly stored dense matrix would require approximately $4.1$ PB [Fig.~\ref{fig:scaling}(c)]. Fast-multipole and matrix-free methods avoid dense storage \cite{corsaro2026}, transition-matrix solvers exploit particle and lattice structure \cite{beutel2024}, and augmented partial factorization recovers many scattering outputs without storing all fields \cite{lin2022}. Bloch periodicity can reduce an infinite metasurface to a small unit-cell problem. The relevant benchmark is therefore an efficient classical solver for the same observable and accuracy, including quantum input construction and fault-tolerant costs \cite{scherer2017}. Register size alone gives no advantage threshold.

Accuracy depends on the optical observable. Interference minima and near-field coupling can amplify coefficient errors. The Supplement derives bounds connecting variational boundary residuals to scattering and Purcell errors, including normalization and source interference \cite{alase2022,bravoprieto2023}. Near the silicon Huygens point, a sufficient worst-case condition for $10\%$ relative accuracy in $p_-$ is a residual below $6.9\times10^{-4}$ in each unscaled dipole system. Bright-lobe agreement gives no such guarantee.

Repeated boundary optimization and readout confirm this sensitivity (shot-based ideal quantum simulator). At approximately $4.5$ million shots, scattering and dark-probability relative root-mean-square errors are $0.38\%$ and $27\%$; at $18$ million shots they are $0.19\%$ and $14\%$. The Supplement gives the full sampling protocol and tolerance tests. Purcell reconstruction also introduces bias through amplitude ratios and quadratic powers. Allocating measurements using known ideal sensitivities reduces the Si total-factor root-mean-square error from $0.0922$ to $0.0167$ at the same approximately $5.0\times10^7$ final-readout shots [Fig.~S3(e)]. This oracle comparison excludes sensitivity-learning costs and optimization noise. Measurement allocation must therefore account for each multipole's contribution to the requested observable.

Even with a prepared state, weak optical probabilities require many measurements. For independent Bernoulli trials, probability $p$ and relative standard error $r$ require approximately
\begin{equation}
 K\simeq\frac{1-p}{pr^2}.
 \label{eq:shots}
\end{equation}
At $1\%$ precision, probabilities $0.5$, $0.01$, and $10^{-4}$ require about $10^4$, $9.9\times10^5$, and $10^8$ accepted trials. The silicon dark probability needs approximately $4.1\times10^5$ accepted trials even at $10\%$ precision [Fig.~\ref{fig:scaling}(d)]. Flag success $0.067$ raises this to roughly $6.1\times10^6$ executions. Direct normalized-state preparation avoids postselection, but $4096$ shots still predict only one dark event. The hardware data show that device errors can exceed this residual signal.

\section*{Conclusions}

The experiments connect electric and magnetic multipole amplitudes to quantum circuits and optical observables. For the known dipole target, raw measured overlaps are $0.998$ on IonQ, $0.993$ on IBM Fez, and $0.793$ on IBM Kingston, versus the ideal unity. For unknown coefficients recovered from boundary measurements, IonQ gives forward directivity $2.984$, about $0.5\%$ below the ideal Huygens value $3$. However, total scattering is $55\%$ below the matched classical silicon result, and the dipole dark-port probability is $0.00529$ versus $2.43\times10^{-4}$. Normalization removes the common amplitude scale, while unequal electric and magnetic amplitudes weaken destructive interference. For an emitter $10$ nm from gold at $510$ nm, IBM Kingston gives total Purcell factor $20.48\,[19.94,21.11]$ versus $20.47$ classically after readout calibration, compared with raw IBM and IonQ estimates $16.03$ and $17.39$. Local calibration addresses qubit-dependent readout errors, while $65\,536$ final shots per circuit reduce sampling noise; the close central value alone does not establish subpercent accuracy.

At the Si inhibition point, raw IonQ gives total factor $2.24$ versus $0.369$ classically and $0.237$ on calibrated IBM. Near-field contractions amplify errors in weak absorptive multipole components, so accurate preparation of a prescribed state does not ensure accurate boundary recovery. These experiments evaluate specific circuits and measurement protocols and do not rank IBM and IonQ or their hardware platforms. The feedback policies, qubit mappings, calibration histories, and shot budgets differ. Accuracy must be established for the chosen physical qubits and calibration state; the reported intervals condition on measured optimizer angles and exclude drift and systematic gate errors. These constraints favor selected optical measurements because full Pauli tomography requires $3^q$ basis settings. Larger nanoantenna arrays, metasurfaces, and other nanophotonic systems require predictions of interacting multipoles and fabrication sensitivity. These single-particle results motivate quantum-assisted design studies. Computational benefit must be assessed against classical solvers at matched accuracy, including coupled-operator construction, state preparation, normalization, and readout costs.

We acknowledge the use of IBM Quantum services and computing credits provided through the IonQ Research Credits Program. A.K. acknowledges support from the U.S. Department of Energy through Brookhaven Science Associates, LLC under Subcontract No. 463609 and Prime Contract No. DE-SC0012704. The conclusions are the author's and do not represent the views of IBM or IonQ.

\nocite{arruda2018,jin2024,majic2020,oskooi2010,ostaszewski2021,stoller2006}
\linespread{0.96}\selectfont
\bibliography{references}
\end{document}